\documentclass[twocolumn,aps,prl,epsf]{revtex4}
\usepackage{epsfig}

\begin{document}

\title{The $\Theta^+$ (1540)
as a heptaquark with the overlap of a pion, a kaon and a nucleon}
\author{P. Bicudo}
\email{bicudo@ist.utl.pt}
\author{G. M. Marques}
\email{gmarques@cfif.ist.utl.pt}
\affiliation{Dep. F\'{\i}sica and CFIF, Instituto Superior T\'ecnico,
Av. Rovisco Pais, 1049-001 Lisboa, Portugal}
\begin{abstract}
We study the very recently discovered $\Theta^+$ (1540) at
SPring-8, at ITEP and at CLAS-Thomas Jefferson Lab. We apply the
same RGM techniques that already explained with success the
repulsive hard core of nucleon-nucleon, kaon-nucleon exotic
scattering, and the attractive hard core present in pion-nucleon
and pion-pion non-exotic scattering. We find that the $K-N$
repulsion excludes the $\Theta^+$ as a $K-N$ s-wave pentaquark. We
explore the $\Theta^+$ as a heptaquark, equivalent to a $N+\pi+K$
borromean bound-state, with positive parity and total isospin
$I=0$. We find that the kaon-nucleon repulsion is cancelled by the
attraction existing both in the pion-nucleon and pion-kaon
channels. Although we are not yet able to bind the total three
body system, we find that the $\Theta^+$ may still be a heptaquark
state. We conclude with predictions that can be tested
experimentally.
\end{abstract}
\maketitle


\par
In this paper we study the exotic hadron $\Theta^+$
\cite{update}
(narrow hadron resonance of 1540 MeV decaying into a $n K^+$) very
recently discovered at Spring-8
\cite{Spring-8},
and confirmed by ITEP
\cite{ITEP}
and by CLAS at the TJNL
\cite{CEBAF}.
This is an extremely exciting state because it may be the first
exotic hadron to be discovered, with quantum numbers that cannot
be interpreted as a quark and an anti-quark meson or as a three
quark baryon. Exotic multiquarks are expected since the early
works of Jaffe
\cite{Jaffe,Strottman},
and some years ago Diakonov, Petrov and Polyakov
\cite{Diakonov}
applied skyrmions to a precise prediction of $\Theta^+$. Very
recent studies suggest that $\Theta^+$ is a pentaquark state
\cite{recent}.
The nature of this particle, its isospin, parity
\cite{Hyodo}
and angular momentum, are yet to be determined.

\par
\begin{figure}[t]
\begin{picture}(100,170)(0,0)
\multiput(133,0)(0,2){76}{$\cdot$}
\put(-69.75,1.75){\epsfig{file=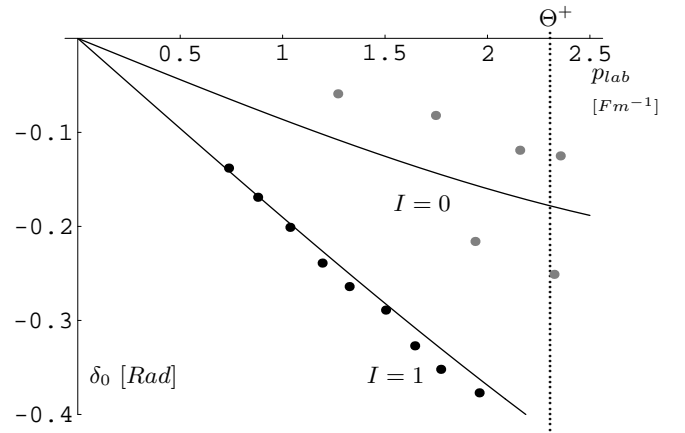,width=8cm}}
\put(150,135){$ p_{lab} $}
\put(150,125){$ _{[Fm^{-1}]} $}
\put(-40,20){$\delta_0  \ [Rad] $}
\put(75,85){$ I=0$}
\put(65,20){$ I=1$}
\put(130,155){$ \Theta^+$}
\end{picture}
\caption{The $I=0$ and $I=1$ experimental
\cite{Barnes}
and theoretical (this paper and ref.
\cite{Bicudo})
s-wave phase shifts as a function of the kaon momentum in the
laboratory frame.} \label{knexp}
\end{figure}

Here we assume a standard Quark Model (QM) Hamiltonian, with a
confining potential and a hyperfine term. Moreover our Hamiltonian
includes a quark-antiquark annihilation term which is the result
of spontaneous chiral symmetry breaking. We start in this paper by
reviewing the QM, and the Resonating Group Method (RGM)
\cite{Wheeler}
which is adequate to study states where several quarks overlap. To
illustrate the RGM, we first apply it to compute the masses of all
the possible s-wave and p-wave $uudd \bar s$ multiquarks. We
verify that the multiquarks computed with the Hamiltonian
(\ref{Hamiltonian}) are too heavy to explain the $\Theta^{+}$
resonance, except for the $I=0$, $J^P=1/2^+$ state. We then apply
the RGM to show that the exotic $N-K$ hard core s-wave interaction
is repulsive
\cite{Bicudo,Bender}.
The result is consistent with the experimental data
\cite{Barnes},
see Fig. \ref{knexp}. We think that this excludes, in our
approach, the $\Theta^+$ as a bare pentaquark $uudd\bar s$ state
or as a tightly bound s-wave $N - K$ narrow resonance. However the
observed mass of $\Theta^+$ is larger than the sum of the $K^+$
and $n$ masses by $1540-940-494=106$ MeV, and this suggests that a
$\pi$ could also be present in this system, in which case the
binding energy would be of the order of $30$ MeV. Moreover this
state of seven quarks would have a positive parity, and would have
to decay to a p-wave $N - K$ system, which is suppressed by
angular momentum, thus explaining the $\Theta^+$ narrow width.
With this natural description in mind we then apply the RGM to
show that $\pi - N$ and $\pi - K$ hard core interactions are
attractive. Finally we put together $\pi - N$, $\pi - K$ and $N-K$
interactions to show that $\Theta^+$ is possibly a borromean
\cite{borromean}
three body s-wave boundstate of a $\pi$, a $N$ and a $K$, with
positive parity and total isospin $I=0$. To conclude, we estimate
the masses of other possible resonances of the $\Theta^+$ family
and compare the QM formalism with other methods that address the
$\Theta^{+}$.

%
%

\par
Our Hamiltonian is the standard QM Hamiltonian,
\begin{equation}
H= \sum_i T_i + \sum_{i<j} V_{ij} +\sum_{i \bar j} A_{i \bar j} \,
\label{Hamiltonian}
\end{equation}
where each quark or antiquark has a kinetic energy $T_i$ with a
constituent quark mass, and the colour dependent two-body
interaction $V_{ij}$ includes the standard QM confining term and a
hyperfine term,
\begin{equation}
V_{ij}= \frac{-3}{16} \vec \lambda_i  \cdot  \vec \lambda_j
\left[V_{conf}(r) + V_{hyp} (r) { \vec S_i } \cdot { \vec S_j }
\right] \ . \label{potential}
\end{equation}
For the purpose of this paper the details of potential
(\ref{potential}) are unimportant, we only need to estimate its
matrix elements. The hadron spectrum is compatible with,
\begin{equation}
\langle V_{hyp} \rangle \simeq \frac{4}{3} \left( M_\Delta-M_N
\right) \label{hyperfine}
\end{equation}
Moreover we include in the Hamiltonian (\ref{Hamiltonian}) a
quark-an\-ti\-quark annihilation potential $A_{i \bar j}$. The
quark-an\-ti\-quark annihilation is constrained when the quark
model produces spontaneous chiral symmetry breaking
\cite{Bicudo3,Bicudo4}.
The annihilation potential $A$ is also present in the $\pi$
Salpeter equation,
\begin{equation}
\left[
\begin{array}{cc}
2 T + V & A \\
A & 2T +V
\end{array}
\right]
\left(
\begin{array}{c}
\phi^+ \\
\phi^-
\end{array}
\right) =
M_\pi
\left(
\begin{array}{c}
\phi^+ \\
-\phi^-
\end{array}
\right)
\label{pion BS}
\end{equation}
where the $\pi$ is the only hadron with a large negative energy
wave-function, $\phi^- \simeq \phi^+$.  In eq. (\ref{pion BS}) the
annihilation potential $A$ cancels most of the kinetic energy and
confining potential $2T+V$. This is the reason why the pion has a very
small mass. From the hadron spectrum and using eq. (\ref{pion BS})
we determine the matrix elements of the annihilation potential,
\begin{eqnarray}
\langle 2T+V \rangle_{S=0} &\simeq& {2 \over 3} (2M_N-M_\Delta)
\nonumber \\
\Rightarrow \langle A \rangle_{S=0} &\simeq&- {2 \over 3} 
(2M_N-M_\Delta)
\ .
\label{sum rules}
\end{eqnarray}
We stress that the QM of eq. (\ref{Hamiltonian}) not only
reproduces the meson and baryon spectrum as quark and antiquark
bound-states (from the heavy quarkonium to the light pion mass),
but it also complies with the PCAC theorems. This includes the
Adler zero
\cite{Bicudo0},
and the Weinberg theorem
\cite{Bicudo0,Bicudo1},
for pion-pion scattering. Therefore our model is adequate to
address the $\Theta^+$, which was predicted by Diakonov, Petrov
and Polyakov in an effective chiral model.

\par
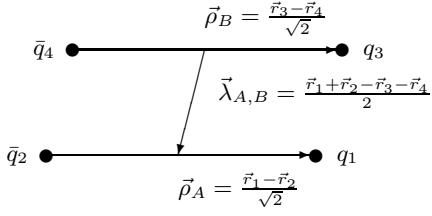
\begin{figure}[t]
%
\begin{picture}(200,80)(0,0)
\put(20,5){
\begin{picture}(120,30)(0,0)
\put(20,0){
\begin{picture}(100,100)(0,0)
\put(0,20){\vector(1,0){100}}
\put(50,5){$\vec \rho_A= {\vec r_1 -\vec r_2 \over \sqrt{2}}$}
\put(60,60){\vector(-1,-4){10}}
\put(65,41){$\vec \lambda_{A,B}={\vec r_1 +\vec r_2 -\vec r_3 -\vec r_4 
\over 2}$}
\put(10,60){\vector(1,0){100}}
\put(60,70){$\vec \rho_{B}={\vec r_3 -\vec r_4 \over \sqrt{2}}$}
\end{picture}}
\put(20,0){
\begin{picture}(100,100)(0,0)
\put(-15,18){$\bar q_2$}
\put(0,20){\circle*{5}}
\put(110,18){$q_1$}
\put(102,20){\circle*{5}}
\put(-5,58){$\bar q_4$}
\put(10,60){\circle*{5}}
\put(120,58){$q_3$}
\put(112,60){\circle*{5}}
\end{picture}}
\end{picture}}
\end{picture}
%
\caption{
Jacobi coordinates of the incoming four quark wave-function
$\phi_A(\rho_A) \, \phi_B(\rho_B) \, \psi(\lambda_{A,B})$. }
\label{RGM coordinates}
\end{figure}
\begin{figure}[t]
\epsfig{file=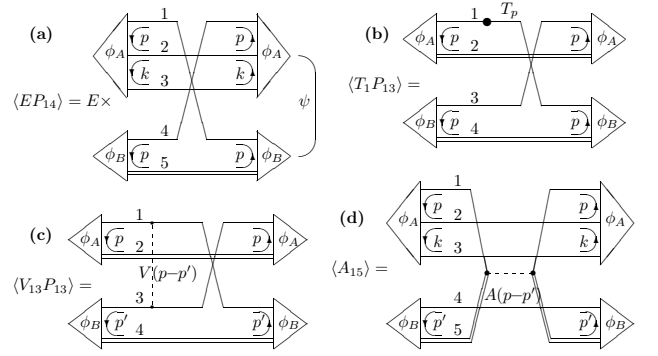,width=8.5cm}
\caption{Examples of RGM overlaps: (a) norm overlap
for meson-baryon interaction; (b) kinetic overlap for meson-meson
interaction; (c) interaction overlap for meson-meson interaction;
(d) annihilation overlap for meson-baryon interaction.}
\label{RGM overlaps}
\end{figure}

With the RGM
\cite{Wheeler}
we compute the matrix elements of the microscopic Hamiltonian
described in eq. (\ref{Hamiltonian}). This method produces both
the energy of a multiquark state and the effective hadron-hadron
interaction. We arrange the wave functions of quarks and
antiquarks in antisymmetrized overlaps of simple colour singlet
quark clusters, the baryons and mesons. This is illustrated in
Fig. \ref{RGM coordinates}, where we show how a tetraquark system
can be arranged in a pair of mesons $A$ and $B$.
%
%
%
Once the internal energies $E_A$ and $E_B$ of the two hadronic
clusters are accounted, the remaining energy of the meson-baryon
or meson-meson system is computed with the overlap of the
inter-cluster microscopic potentials,
\begin{eqnarray}
V_{\text{bar } A \atop \text{mes } B}&=& \langle \phi_B \, \phi_A |
-( V_{14}+V_{15}+2V_{24}+2V_{25} )3 P_{14}
\nonumber \\
&& +3A_{15} | \phi_A \phi_B \rangle /
\langle \phi_B \, \phi_A | 1- 3 P_{14} | \phi_A \phi_B \rangle
\nonumber \\
V_{\text{mes } A \atop \text{mes } B}&=& \langle \phi_B \,
\phi_A | (1+P_{AB})[ -( V_{13}+V_{23}+V_{14}+V_{24})
\nonumber \\
&& \times P_{13} +A_{23} ]| \phi_A \phi_B \rangle
\ ,
\label{eq:overlapkernel}
\end{eqnarray}
where $P_{ij}$ stands for the exchange of particle $i$ with
particle $j$, see Fig. \ref{RGM overlaps}. It is clear that quark
exchange is necessary in the inter-cluster matrix elements, to
provide colour octets to match the Gell-Mann matrices $\lambda_i$.
In what concerns the quark-quark(antiquark) potential $V_{ij}$,
the spin independent part of the interaction vanishes because the
clusters are colour singlets. The only potential which may
contribute is the hyperfine potential. For instance when a cluster
is a spin singlet, the minus phase from colour is reverted by a
minus phase from spin, and it turns out that the inter-cluster
potentials result in a positive energy shift. As for annihilation,
it only occurs in non-exotic channels. We therefore find that the
total energy of the multiquark system is the sum of the masses of
the clusters $E_A+E_B$, plus the result of to eq.
(\ref{hyperfine}) or eq. (\ref{sum rules}) times an algebraic
colour $\times$ spin $\times$ flavour factor and a geometric
spatial overlap
\cite{Ribeiro3}.
%
%

\par
We now apply the RGM to compute the mass of all the possible
s-wave and p-wave $uudd \bar s $ multiquarks. The p-wave states
were suggested by Diakonov, Petrov and Polyakov
\cite{Diakonov} and by effective pion exchange models
\cite{Robson,Stancu}.
The $I=2$ flavour case was suggested by Capstick, Page and Roberts
\cite{Page},
as a candidate to explain the narrow width of $\Theta^+$. Using
the RGM we find that the multiquark masses are essentially the sum
of $N$, $N^*$ or $\Delta$, and $K$ or $K^*$ masses, further
shifted by the hyperfine and annihilation interactions. Thus the
multiquark mass results in a sum rule which is directly evaluated
from the hadron spectrum. After computing the different multiquark
masses, we find that the lightest multiquarks have the quantum
numbers $I=0$, $J^P= 1/2^-$ with a mass of 1530 MeV, and $I=1, \,
J^P= 1/2^-$ with a mass of 1775 MeV. All the other multiquarks,
with higher spin, isospin, or angular momentum quantum numbers
have a mass at least 400 MeV heavier than $M_N+M_K$. This is
expected with our standard quark model Hamiltonian defined in eq.
(\ref{Hamiltonian}). The positive mass shifts of more than 400 MeV
are either produced by the hyperfine interaction estimated in eq.
(\ref{hyperfine}), or by a p-wave excitation which is also
directly estimated from the baryon spectrum.

\par
We now proceed to study the $\Theta^+$ in hadron-hadron
scattering, in particular we address the most favorable $I=0$,
$J^P= 1/2^-$ channel which is coupled to $K-N$ s-wave scattering.
The RGM was used by Ribeiro
\cite{Ribeiro}
to show that in exotic hadron-hadron scattering, the quark-quark
potential together with Pauli repulsion of quarks produces a
repulsive short range interaction. For instance this explains the
$N - N$ hard core repulsion
\cite{Bicudo2},
preventing nuclear collapse. Deus and Ribeiro
\cite{Deus}
used the same RGM to show that, in non-exotic channels, the
quark-antiquark annihilation could produce a short core
attraction. Recently the short core attraction was further
understood
\cite{Bicudo0,Bicudo1},
and it was shown that the QM and the RGM reproduce the Weinberg
Theorem for $\pi-\pi$ scattering, with attraction in the $I=0$
channel and repulsion in the exotic $I=2$ channel. A convenient
basis for the wave-functions is the harmonic oscillator basis,
\begin{equation}
\phi_{000}^\alpha(p_\rho) =  {\cal N_\alpha}^{-1}
\exp\left({ - {{p_\rho}^2 \over 2 \alpha ^2}}\right) \ , \ \
{\cal N_\alpha} = \left({ \alpha \over 2 \sqrt{\pi}}\right)^{3 \over 2} 
\ ,
\label{basis}
\end{equation}
where, in the case of vanishing external momenta $p_A$ and $p_B$,
the momentum integral in eq. (\ref{eq:overlapkernel}) is simply
${\cal N_\alpha}^{-2}$. For instance in the matrix element of the
exchange operator $P_{13}$ in a meson-meson system, the
coordinates of the incoming $\phi_A\phi_B$ functions are
illustrated in Fig. \ref{RGM coordinates}, while the coordinates
of the outgoing $\phi_A^\dagger\phi_B^\dagger$ have the quark 1
and 3 exchanged. We summarize
\cite{Bicudo,Bicudo1,Bicudo2}
the effective potentials computed for the different channels,
\begin{eqnarray}
V_{K-N}&=&\frac{4}{3}(M_\Delta-M_N ) \frac{{1\over 2}+{1 \over 3}  \vec 
\tau_K \cdot \vec
\tau_N}{\frac{3}{4}-\frac{1}{3}  \vec \tau_K \cdot \vec \tau_N}
{\cal N_\alpha}^{-2} \ ,
\nonumber \\
V_{\pi-N}&=& \frac{2}{9}( 2M_N - M_\Delta ) \,  \vec \tau_\pi \cdot
\vec \tau_N \, {\cal N_\alpha}^{-2} \ ,
\nonumber \\
V_{\pi-K}&=& \frac{8}{27}(2 M_N - M_\Delta) \, \vec \tau_\pi \cdot
\vec \tau_K \, {\cal N_\alpha}^{-2} \ ,
\label{zero p}
\end{eqnarray}
where $\vec \tau$ are the isospin matrices. The results of eqs.
(\ref{hyperfine}) and (\ref{sum rules}) are recognised in eq.
(\ref{zero p}). The vanishing momentum case of eq. (\ref{zero p})
is sufficient to compute the scattering lengths with the Born
approximation, and to estimate that $\alpha \simeq 3$ Fm$^{-1}$.
The estimation of $\alpha$ is an important by-product of this
method because the hadronic size can not be estimated directly by
the hadronic charge radius which is masked by the vector meson
dominance.

\par
However we are interested in binding and therefore we have to
proceed to the finite momentum case. Then the effective potentials
in eq. (\ref{zero p}) turn out to be multiplied by the gaussian
separable factor, $\exp\left[- \frac{{p_\lambda}^2}{2\beta^2
}\right] \int \frac{d^3 p'_\lambda}{(2 \pi)^3} \exp\left[- \frac{
{p'_\lambda}^2}{2 \beta^2}\right]$. In the exotic $K - N$
channels, we can prove this result and we also show that the new
parameter $\beta=\alpha$. This occurs because the overlaps
decrease when the relative momentum of hadrons $A$ and $B$
increases. In the non-exotic $\pi - N$ and $\pi - K$ channels the
present state of the art of the RGM does not allow a precise
determination of the finite momentum overlap. In this case $\beta$
is a new degree of freedom that will be fitted with the
experiment. In particular, in pionic channels the Adler zero
\cite{Bicudo0}
disappears at finite momentum, thus contributing to the
enhancement of the overlap when momentum increases. Nevertheless,
we expect that eventually the overlap decreases due to the
geometric wave-function overlap in momentum space. So we expect
that $\beta >>\alpha$ in the pionic channels. This
parameterization in a separable potential enables us to use
standard techniques
\cite{Bicudo}
to exactly compute the scattering $T$ matrix. The scattering
length $d \delta_0 / dk_{c.m.}$ is,
\begin{equation}
a=-{ \sqrt{\pi} \over \alpha} { 4 \mu \, v \over \alpha ^2 +
{\beta \over \alpha} 4 \,  \mu \, v } \ .
\label{teor scatt}
\end{equation}
The parameters and results for the relevant channels are
summarized and compared with experiment
\cite{Barnes,Itzykson,Nemenov}
in Table \ref{scattering lengths}. We have fitted $\alpha$ with
the $I=1$ $K-N$ scattering. We use $\beta=\alpha$ in the repulsive
channels. We fit $\beta$ with the appropriate scattering lengths
in the attractive $I=1/2$ pionic channels.
%
%
\begin{table}[t]
\begin{tabular}{c|cccccc}
channel                   & $\mu$  & $v_{th}$&$\alpha$ &$\beta$& 
$a_{th}$& $a _{exp}$ \\
\hline
$  K-N_{ I=0 }         $ & $ 1.65$ & $ 0.50$ & $ 3.2$ & $ 3.2$ & 
$-0.14$ & $ -0.13\pm 0.04 $
\cite{Barnes} \\
$  K-N_{ I=1 }         $ & $ 1.65$ & $ 1.75$ & $ 3.2$ & $ 3.2$ & 
$-0.30$ & $ -0.31\pm 0.01 $
\cite{Barnes} \\
$ \pi-N_{I={1\over 2}} $ & $ 0.61$ & $-0.73$ & $ 3.2$ & $ 11.4$ & $ 
0.25$ & $  0.246\pm 0.007$
\cite{Itzykson} \\
$ \pi-N_{I={3\over 2}} $ & $ 0.61$ & $ 0.36$ & $ 3.2$ & $ 3.2$ & 
$-0.05$ & $ -0.127\pm 0.006$
\cite{Itzykson} \\
$ \pi-K_{I={1\over 2}} $ & $ 0.55$ & $-0.97$ & $ 3.2$ & $ 10.3$ & $ 
0.35$ & $  0.27\pm 0.08 $
\cite{Nemenov} \\
$ \pi-K_{I={3\over 2}} $ & $ 0.55$ & $ 0.49$ & $ 3.2$ & $ 3.2$ & 
$-0.06$ & $ -0.13\pm 0.06 $
\cite{Nemenov} \\
\hline
\end{tabular}
\caption{ This table summarizes the parameters $\mu , \, v \,
,\alpha \, , \beta$ (in Fm$^{-1}$)
 and scattering lengths $a$ (in Fm) .}
\label{scattering lengths}
\end{table}

\par
Let us first apply our method to the $K-N$ exotic scattering. The
only channel where we find a pentaquark with a mass close to 1540
MeV is the $I=0$, $J^P=1/2^+$ channel. This pentaquark is open to
the $K-N$ s-wave channel, which is repulsive for $I=0$, therefore
its decay width is quite large. We think that this excludes
$\Theta^+$ as a $uudd\bar s$ pentaquark, because its experimental
decay width is quite narrow. With our method we reproduce the
$K-N$ exotic s-wave phase shifts, see Fig. \ref{knexp}, where
indeed there is no evidence for the $\Theta^+$ state. In what
concerns the $\pi - N$ system and the $\pi - K$ systems, the
corresponding parameters in Table \ref{scattering lengths} are
almost identical. There we find repulsion for $I=3/2$ and
attraction for $I=1/2$. The repulsion in the $I=3/2$ channel
prevents a bound state in this channel. In what concerns the
$I=1/2$ channel, the attraction is not sufficient to provide for a
bound state, because the $\pi$ is quite light and the attractive
potential is narrow. From eq. (\ref{teor scatt}) we conclude that
we have binding if the reduced mass is $\mu \geq -{\alpha^3 \over
4 \beta v }$. With the present parameters this limit is $ 0.86
\rightarrow 1.07$ Fm$^{-1}$. This is larger than the $\pi$ mass,
therefore it is not possible to bind the $\pi$ to a $K$ or to a
$N$. All that we can get is a very broad resonance. For instance
in the $\pi - K$ channel this is the kappa resonance
\cite{Rupp1}, which has been recently confirmed by the scientific
community. However, with a doubling of the interaction, produced
by a $K$ and a $N$, we expect the $\pi$ to bind.

\par
\begin{figure}[t]
%
\begin{picture}(200,70)(0,0)
\put(20,5){
\begin{picture}(120,70)(0,0)
\put(20,0){
\begin{picture}(100,100)(0,0)
\put(0,0){\oval(30,30)}
\put(0,5){$N$}
\put(-12,-8){$I={1\over 2}$}
\put(40,55){\oval(30,30)}
\put(35,43){$\pi$}
\put(30,58){$I=1$}
\put(80,0){\oval(30,30)}
\put(70,5){$K$}
\put(68,-8){$I={1\over 2}$}
\end{picture}}
\put(20,0){
\begin{picture}(20,100)(0,0)
\put(40,10){\oval(90,14)}
\put(25,5){$I=1$}
\put(41,40){\oval(20,20)[tl]}
\put(31,40){\line(1,-1){40}}
\put(41,50){\line(1,-1){40}}
\put(71,10){\oval(20,20)[br]}
\put(43,30){$I$}
\put(46,27){$=$}
\put(54,25){$1 \over 2$}
\put(6,10){\oval(20,20)[bl]}
\put(-4,10){\line(1,1){40}}
\put(5,-1){\line(1,1){40}}
\put(36,40){\oval(20,20)[tr]}
\put(15,21){$I$}
\put(20,26){$=$}
\put(26,29){$1 \over 2$}
\end{picture}}
\put(20,0){
\begin{picture}(20,100)(0,0)
\put(-20,60){Total}
\put(-20,50){$I=0$}
\end{picture}}
\end{picture}}
\end{picture}
%
\caption{ The isospin couplings in the $Z/\Theta$. } \label{ISO 1540}
\label{isospin}
\end{figure}
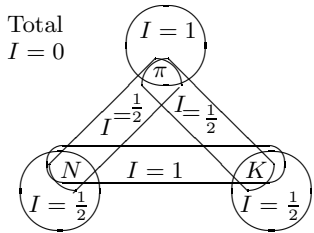

We now investigate the borromean \cite{borromean} binding of the
exotic $\Theta^+$ constituted by a $N$, $K$ and $\pi$ triplet. In
what concerns isospin, we need the $\pi$ to couple to both $N$ and
$K$ in $I=1/2$ states for attraction. It turns out that the
possible $K-N$ states decaying in the observed $K^+-n$ are
obtained with a linear combination of one total $I=0$ and two
total $I=1$ states. Because the Hamiltonian does not mix the total
$I=0$ with the total $I=1$, we can study them separately. The
$I=0$ state includes both the $\pi-N$ and $\pi-K$ in $I=1/2$
states, however the $N-K$ pair is in a $I=1$ state, nevertheless
we expect that this is not sufficiently repulsive to cancel the
binding. In what concerns the $I=1$ states, they both have a
significant $I=3/2$ either in the $\pi - N$ pair or in the $\pi -
K$ pair. This excludes the $I=1$ states because either the $N$ or
the $K$ would not be bound. The only real candidate for binding is
the total $I=0$ state, see Fig. \ref{isospin}.
%
%
%

\par
\begin{figure}[t]
\begin{picture}(100,120)(0,0)
\put(-70,0){\epsfig{file=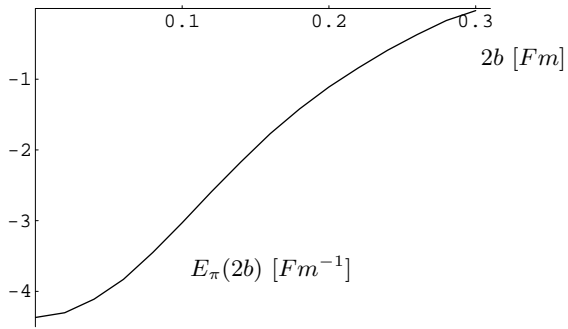,width=6.5cm}}
\put(110,100){$ 2b \ [Fm] $}
\put(0,20){$E_\pi (2b) \ [Fm^{-1}] $}
\end{picture}
\caption{ $\pi$ energy as a function of the
coordinate $ |\vec r_N-\vec r_K|=2b$.}
\label{pion energy}
\end{figure}

Since the $\pi$ is much lighter that the $N - K$ system, we study
the borromean binding adiabatically. As a first step, we start by
assuming that the $K$ and $N$ are essentially motionless and
separated by $\vec r_N-\vec r_K= 2 \vec b$. This will be
improved later. We also take advantage of the similarities in
Table \ref{scattering lengths} to assume that the two heavier
partners of the $\pi$ have a similar mass of 3.64 Fm$^{-1}$ and
interact with the $\pi$ with the same separable potential. Then 
we solve the bound state equation for a $\pi$ in the potential
$V_{\pi-N}+ V_{\pi-K}$,
\begin{equation}
v \, \frac{\phi_{000}^\beta(\vec r_\pi- \vec b)}{{\cal N}_\alpha}
\int d^3 r'_\pi  \frac{\phi_{000}^\beta(\vec r'_\pi- \vec b)}{
{\cal N}_\alpha} + (-\vec b \leftrightarrow \vec b)
\end{equation}
where this potential is wider in the direction of the $K - N$
axis. Nevertheless this potential remains separable and we can
apply standard techniques to find the poles of the $T$ matrix,
which occur exactly at the $\pi$ energy. The $\pi$ energy is
depicted if Fig. \ref{pion energy}, and it is negative as
expected.

\par
Once the $\pi$ binding energy is determined, we include it in the
potential energy of the $K - N$ system, which becomes the sum of
the repulsive $K - N$ potential and the $\pi$ energy. We find that
for short distances the total potential is indeed attractive.
Finally, using this $K - N$ potential energy we solve the
Schr\"odinger equation for the system, thus including the
previously neglected $K$ and $N$ kinetic energies. However here we
are not able to bind the $K-N$ system, because the total effective
$K-N$ potential is not sufficiently attractive to cancel the
positive $K-N$ kinetic energy. Nevertheless the $K$ is heavier
than the $\pi$, thus a small enhancement of the attraction would
suffice to bind the heptaquark. We remark that existing examples
of narrow resonances with a trapped $K$ are the $f_0(980)$
\cite{Rupp1},
the $D_s(2320)$
\cite{Rupp2}
very recently discovered at BABAR
\cite{Babar},
and possibly the $\Lambda(1405)$. Moreover we expect that meson
exchange interactions and the irreducible three-body overlap of
the three hadrons, that we did not include here, would further
increase the attractive potential. Therefore it is plausible that
a complete computation will eventually bind the $K-\pi-N$ system.

\par
To conclude, in this paper we address the $\Theta^+$ hadron with a
standard quark model Hamiltonian, where the quark-antiquark
annihilation is constrained by the spontaneous breaking of chiral
symmetry. We first find that the $\Theta^+$ hadron very recently
discovered cannot be an s-wave or p-wave pentaquark. To provide
sufficient binding, it would be necessary to change the
spin-independent interaction, for instance to a bag model
potential
\cite{Jaffe},
or to change the hyperfine interaction to an effective pion
exchange interaction
\cite{Robson,Stancu}.
%

\par
We also find that the $\Theta^+$ may be essentially a heptaquark
state, composed by the overlap of a $\pi$, a $K$ and a $N$. This
scenario has many interesting features. The $\Theta^+$ would be,
so far, the only hadron with a trapped $\pi$. Moreover the $\pi$
would be trapped by a rare three body borromean effect. And the
decay rate to a $K$ and a $N$ would be suppressed since the $\pi$
needs to be absorbed with a derivative coupling, while the
involved hadrons have a very low momentum in this  $\Theta^+$
state. Because the $\Theta^+$ would be composed by a $N$ and two
pseudoscalar mesons, its parity would be positive,
$J^P={1\over2}^+$. Our result with a positive parity agrees with
the prediction of Diakonov, Petrov and Polyakov, although these
authors suggested that parity was due to a p-wave excitation. The
isospin of the $\Theta^+$ would be $I=0$ in order to ensure the
attraction of the $\pi$ both by the $N$ and the $K$.

\par
We now detail other possible three body hadronic molecules that
can be searched experimentally. Our mechanism is also expected to
bind either a $K \pi K$ or a $N \pi N$ into a $I=0$ narrow resonance
with negative parity, slightly below threshold. Because the
antiquark $\bar s$ is just a spectator in our scenario, similar
states with flavour $uudd \bar c$ or $ uudd \bar b$ are also
expected. Moreover we estimate in a three meson scenario the
masses of the positive parity anti-decuplet predicted by Diakonov,
Petrov and Polyakov. A three hadron molecule $N K \bar K$ is
expected in the iso-doublet including the state $uud$, with a mass
close to 1900 MeV. In the iso-triplet including the state $uus$,
the binding is comparable with the $\Theta^+$ case and a $N \bar K
\pi $ is suggested, again at a mass close to 1540 MeV. The
iso-quadruplet including the state $uuss\bar d$ is also expected
to include a three body $N \bar K \bar K$ binding with a mass
close to 1900 MeV. Therefore an anti-decuplet is possible in the
three hadron scenario, but the masses differ
\cite{Close}
from the Diakonov, Petrov and Polyakov prediction
\cite{Diakonov}.

\par
We plan to proceed with the challenging study of the heptaquark
state, including the p-wave $K-N$ coupled channel, meson exchange
interactions and three-body potentials. This will be done
elsewhere.
%
%
\begin{table}[t]
\begin{tabular}{c|cccc}
flavour    & $N K \pi_{I=0}$ & $N K \bar K_{I=1/2}$ & $N \bar K 
\pi_{I=3/2}$ & $N \bar K \bar K_{I=2}$ \\
\hline
mass [MeV]    & 1540  &  1900 & 1540 & 1900
\\
\hline
\end{tabular}
\caption{ Approximate masses of the anti-decuplet .}
\label{antidecuplet}
\end{table}

\acknowledgements
We are very grateful to George Rupp for pointing our attention to
the pentaquark state.

The work of G. Marques is supported by Funda\c c\~ao para a
Ci\^encia e a Tecnologia under the grant SFRH/BD/984/2000.



\begin{thebibliography}{00}
%
\bibitem{update}
For a recent online review see P. Schewe, J. Riordon, B. Stein;
Physics News Update {\bf 644},  1,  June 30 (2003).
%
\bibitem{Spring-8}
T. Nakano et al, arXiv:hep-ex/0301020, submitted to
Phys. Rev. Lett.
%
\bibitem{ITEP}
V. Barmin et al, arXiv:hep-ex/0304040.
%
\bibitem{CEBAF}
K. Hicks, to be published in the proceedings of the Conference on the 
Intersections
of Particle and Nuclear Physics, New York may 2003, edited by Zohreh 
Parsa (2003).
%
\bibitem{Jaffe}
R.~L.~Jaffe,
Phys.\ Rev.\ D {\bf 15} 281 (1977).
%
\bibitem{Strottman}
D.~Strottman,
Phys.\ Rev.\ D {\bf 20} 748 (1979).
%
\bibitem{Diakonov}
D.~Diakonov, V.~Petrov and M.~V.~Polyakov,
Z.\ Phys.\ A {\bf 359} 305 (1997) [arXiv:hep-ph/9703373].
%
%
\bibitem{recent}
%
H.~Walliser and V.B.~Kopeliovich
arXiv:hep-ph/0304058;
%
H.~Gao and B.-Q.~Ma,
Mod.\ Phys.\ Lett.\ A {\bf 14} 2313 (1999)
[arXiv:hep-ph/0305294];
%
B.~Wybourne
arXiv:hep-ph/0307170;
%
A.~Hosaka
arXiv:hep-ph/0307232;
%
M.~Karliner and H.~Lipkin
arXiv:hep-ph/0307243;
%
M.~Karliner and H.~Lipkin
arXiv:hep-ph/0307343;
%
S.~L.~Zhu,
arXiv:hep-ph/0307345;
%
Xun Chen, Yajun Mao and Bo-Qiang Ma
arXiv:hep-ph/0307381;
%
C.~E.~Carlson, C.~D.~Carone, H.~J.~Kwee and V.~Nazaryan,
arXiv:hep-ph/0307396;
%
L.~Chen, V.~Greco, C.~Ko, S.~Lee and W.~Liu
nucl-th/0308006.
%
%
\bibitem{Hyodo}
T.~Hyodo, A.~Hosaka and E.~Oset,
arXiv:nucl-th/0307105.
%
\bibitem{Wheeler}
J.~Wheeler, Phys.\ Rev.\ {\bf 52}, 1083 (1937); {\bf 52}, 1107
(1937).
%
\bibitem{Bicudo}
P.~Bicudo and J.~E.~Ribeiro,
Z.\ Phys.\ C {\bf 38}, 453 (1988); 
P.~Bicudo, J.~E.~Ribeiro and J.~Rodrigues,
Phys.\ Rev.\ C {\bf 52}, 2144 (1995). 
%
\bibitem{Bender}
I.~Bender, H.~G.~Dosch, H.~J.~Pirner and H.~G.~Kruse,
Nucl.\ Phys.\ A {\bf 414}, 359 (1984). 
%
\bibitem{Barnes}
T.~Barnes and E.~Swanson, Phys.\ Rev.\ C {\bf 49} 1166 (1994);
J.~S.~Hyslop, R.~A.~Arndt, L.~D.~Roper and R.~L.~Workman,
Phys.\ Rev.\ D {\bf 46} 961 (1992).
%
\bibitem{borromean}
M.~V.~Zhukov, B.~V.~Danilin, D.~V.~Fedorov, J.~M.~Bang, I.~J.~Thompson 
and J.~S.~Vaagen,
Phys.\ Rept.\ {\bf 231}, 151 (1993);
J.~M.~Richard,
arXiv:nucl-th/0305076. 
%
\bibitem{Bicudo3}
P.~Bicudo and J.~E.~Ribeiro,
Phys.\ Rev.\ D {\bf 42}, 1611 (1990); 
1625 (1990); 
1635 (1990). 
%
\bibitem{Bicudo4}
P.~Bicudo,
Phys.\ Rev.\ C {\bf 60}, 035209 (1999). 
%
\bibitem{Bicudo0}
P.~Bicudo, 
Phys.\ Rev.\ C {\bf 67}, 035201 (2003). 
%
\bibitem{Bicudo1}
P.~Bicudo, S.~Cotanch, F.~Llanes-Estrada, P.~Maris, J.~E.~Ribeiro and
A.~Szczepaniak,
Phys.\ Rev.\ D {\bf 65}, 076008 (2002) [arXiv:hep-ph/0112015].
%
\bibitem{Ribeiro3}
J. E. Ribeiro, Phys. Rev. D {\bf 25}, 2406 (1982);
E. van Beveren, Zeit. Phys. C {\bf 17}, 135 (1982).
%
\bibitem{Robson}
D. Robson, {\em in} proceedings of the Topical Conference on Nuclear 
Chromodynamics,
Argonne 1988, Eds. J. Qiu and D. Sivers (World Scientific), 174 
(1988).
%
\bibitem{Stancu}
F.~Stancu and D.~O.~Riska,
Phys.\ Lett.\ B {\bf 575}, 242 (2003)
[arXiv:hep-ph/0307010].
%
\bibitem{Page}
S.~Capstick, P.~R.~Page and W.~Roberts,
Phys.\ Lett.\ B {\bf 570}, 185 (2003)
[arXiv:hep-ph/0307019].
%
\bibitem{Ribeiro}
J.~E.~Ribeiro, Z.\ Phys.\ C {\bf 5}, 27 (1980).
%
\bibitem{Bicudo2}
P.~Bicudo, M.~Faria, G.~M.~Marques and J.~E.~Ribeiro,
arXiv:nucl-th/0106071. 
%
\bibitem{Deus}
J.~Dias de Deus and J.~E.~Ribeiro Phys.\ Rev.\ D {\bf 21}, 1251
(1980).
%
\bibitem{Itzykson}
C.~Itzykson and J.~B.~Zuber, ``Quantum Field Theory,'' published by
Mcgraw-hill,
 New York, Usa (1980).
%
\bibitem{Nemenov}
L. Nemenov, ``Lifetime Measurement of $\pi^+\pi^-$ and $\pi^\pm
K^\pm$ atoms to test low energy QCD'' L23 Letter of Intent for
Nuclear and Particle Physics Experiments at the J-PARC (2003).
%
\bibitem{Rupp1}
E. van Beveren, C. Dullemond, C. Metzger, J. E. Ribeiro, T.A. Rijken, and 
G. Rupp,
Z. Phys. C 30, 615 (1986).
%
\bibitem{Rupp2}
E. van Beveren and G. Rupp,  Phys.Rev.Lett.{\bf 91}, 012003(2003)
 [arXiv:hep-ph/0305035].
%
\bibitem{Babar}
B.~Aubert {\it et al.} [BABAR Collaboration],
Phys.\ Rev.\ Lett.\ {\bf 90}, 242001 (2003) [arXiv:hep-ex/0304021].
%
\bibitem{Close}
F.~ Close  {\em in} the closing talk of
Hadron2003, Aschaffenburg, Germany (2003).
\end{thebibliography}
\end{document}